\documentclass[aip,amsmath,amsthm,amssymb,citeautoscript,reprint]{revtex4-2}
\usepackage{graphicx, float, xcolor, pgf} 
\usepackage{physics, siunitx}
\usepackage[caption=false, justification=centerlast]{subfig}

\usepackage{hyperref}
\hypersetup{
    pdfauthor={Amartya Bose},
    pdftitle={Adaptive Kink Filtration: Achieving Practical Size-Independence of Path Integral Simulations},
    colorlinks=true,
}

\allowdisplaybreaks

\begin{document}
\title{Adaptive Kink Filtration: Achieving Asymptotic Size-Independence of Path Integral Simulations Utilizing the Locality of Interactions}
\author{Amartya Bose}
\affiliation{Department of Chemical Sciences, Tata Institute of Fundamental Research, Mumbai 400005, India}
\email{amartya.bose@tifr.res.in}
\begin{abstract}
    Recent method developments involving path integral simulations have come a
    long way in making these techniques practical for studying condensed phase
    non-equilibrium phenomena. One of the main difficulties that still needs to
    be surmounted is the scaling of the algorithms with the system
    dimensionality. The majority of recent techniques have only changed the
    order of this scaling (going from exponential to possibly a very high
    ordered polynomial) and not eased the dependence on the system size. In this
    current work, we introduce an adaptive kink filtration technique for path
    generation approach that leverages the locality of the interactions present
    in the system and the consequent sparsity of the propagator matrix to remove
    the asymptotic size dependence of the simulations for the propagation of
    reduced density matrices. This enables the simulation of larger systems at a
    significantly reduced cost. This technique can be used both for simulation
    of non-equilibrium dynamics and for equilibrium correlation functions, and
    is demonstrated here using examples from both --- simulating the excitonic
    dynamics in bacteriochlorophyll chains and their absorption and emission
    spectra. We show that the cost becomes constant with the dimensionality of
    the system. The only place where a system size-dependence still remains is
    the calculation of the dynamical maps or propagators which are important for
    the transfer tensor method. The cost of calculating this
    solvent-renormalized propagator is the same as the cost of propagating all
    the elements of the reduced density matrix, which scales as the square of
    the size. This adaptive kink-filtration technique promises to be
    instrumental in extending the affordability of path integral simulations for
    very large systems.
\end{abstract}
\maketitle

\section{Introduction}
Simulation of the dynamics of open quantum systems has been a perpetual
challenge owing to the exponential growth of computational complexity of
quantum dynamics with the number of degrees of freedom. Avoiding this ``curse of
dimensionality'' and side-stepping the exponential scaling has become a
cornerstone of method development in the field. The relevance of such efforts
are amply illustrated by the relevance of these simulations in the world of
artificial photosynthesis and quantum computing. While wave function propagation
approaches like density matrix renormalization group
(DMRG)~\cite{whiteDensityMatrixFormulation1992} and its time-dependent variant
(t-DMRG)~\cite{whiteRealTimeEvolutionUsing2004,
paeckelTimeevolutionMethodsMatrixproduct2019}, and the (multilevel-)
multiconfiguration time-dependent Hartree
((ML-)MCTDH)~\cite{meyerMulticonfigurationalTimedependentHartree1990,
wangMultilayerFormulationMulticonfiguration2003,
wangMultilayerMulticonfigurationTimeDependent2015} are capable of handling large
systems, they are still not able to handle a continuum of environment degrees of
freedom populated at finite temperatures efficiently.

A lucrative alternative to wave function-based methods are methods that simulate
the dynamics of the reduced density matrix. While approximate methods like
Bloch-Redfield master equation, Lindblad master equation, and multichromophoric
incoherent F\"orster theory are commonly used in different circumstances, path
integrals offer a way to simulate the dynamics in a numerically rigorous manner
incorporating the effects of the dissipative media without any approximation
using the Feynman-Vernon influence
functional~\cite{feynmanTheoryGeneralQuantum1963}. The quasi-adiabatic
propagator path integral~\cite{makriTensorPropagatorIterativeI1995,
makriTensorPropagatorIterativeII1995, makriLongtimeQuantumSimulation1996}
(QuAPI) family of methods and the hierarchical equations of
motion~\cite{tanimuraQuantumClassicalFokkerPlanck1991,
tanimuraNumericallyExactApproach2020} (HEOM) family of methods form two of the
most commonly used frameworks that allow for such calculations. HEOM has
historically been limited to using a Drude-Lorentz description of the
dissipative environment. A flurry of recent work though has been geared towards
expanding the capabilities of the original algorithm to dealing with
non-Drude-Lorentz baths and especially more structured
environments~\cite{rahmanChebyshevHierarchicalEquations2019,
ikedaGeneralizationHierarchicalEquations2020, yanNewMethodImprove2020,
yanEfficientPropagationHierarchical2021, xuTamingQuantumNoise2022,
keTreeTensorNetwork2023}.

While these numerically rigorous path integral methods are capable of giving the
correct dynamics irrespective of the coupling involved, the cost of the
calculations grow exponentially with the system size within the non-Markovian
memory length. The transfer tensor method
(TTM)~\cite{cerrilloNonMarkovianDynamicalMaps2014} and the small matrix
decomposition of path integrals
(SMatPI)~\cite{makriSmallMatrixDisentanglement2020} alleviate the cost of
propagating the reduced density matrix beyond the memory length. A different set
of methods have been aimed at addressing the simulation within the memory
length, which can then be leveraged by the previous two methods to simulate
beyond it. This intrinsically non-Markovian portion of the dynamics proves to be
more challenging to handle. While the blip
decomposition~\cite{makriBlipDecompositionPath2014,
makriIterativeBlipsummedPath2017} was used to bring down the cost based on the
structure of the influence functional, various methods based on tensor
networks~\cite{strathearnEfficientNonMarkovianQuantum2018,
jorgensenExploitingCausalTensor2019, bosePairwiseConnectedTensor2022,
boseQuantumCorrelationFunctions2023} utilize the relative locality of the
non-Markovian memory to improve the scaling of the full path simulations.
Furthermore, the multi-site tensor network path integral
approach~\cite{boseMultisiteDecompositionTensor2022} has used tensor networks to
capitalize the short-ranged nature of the interactions between various units of
a long aggregate to improve the performance of path integrals for extended
systems. These methods have made the application of path integrals to large
molecular aggregates increasingly approachable~\cite{boseTensorNetworkPath2022}.
All of these tensor network-based methods bring down the exponential scaling of
the algorithm to some polynomial scaling.

However, one of the important problems that remain is how to tackle larger
systems. The cost of the full-path simulation grows extremely rapidly with the
system size. Recently, Makri has shown that limiting the number of kinks (or
time-steps where the state of the system changes in a given path) allowed in the
paths proves to be an interesting alternative approach to building the path list
for the simulations~\cite{makriKinkSumLongMemory2024}. While it is also able to
curb the exponential scaling to a polynomial algorithm, this kink-summed path
integral approach still scales as a high-degree polynomial of the size of the
system. Here, it should be mentioned that the modular path integral
(MPI)~\cite{makriModularPathIntegral2018} and the multi-site tensor network path
integral (MS-TNPI)~\cite{boseMultisiteDecompositionTensor2022} are capable of
handling large systems with relatively local interactions.

In this work, we ask if it is possible to reduce the complexity of path
integrals for extremely large systems. The dream is to attempt to ensure that
for very large systems, asymptotically, the scaling of the path integral becomes
a constant instead of depending on the system size. A crucial step towards this
goal is in identification of the fact that even in presence of long-ranged
interactions, for all physical systems, the extent of these long-ranged
interactions are much smaller than size of the aggregates. As a result, the
system propagators become sparse. This current work leverages the sparsity of
the propagator matrix in developing a path integral filtration technique in
which the number of paths does not scale with the system size beyond some
threshold size. The kink filtration technique developed here additionally does
not depend on such a representation or particular structures of the Hamiltonian.
This paper is organized as follows: in Sec.~\ref{sec:motivation} we motivate the
method by exploring some simple closed-system problems. The path generation
algorithm is developed using these intuitions. Then in
Sec.~\ref{sec:numerical-examples} we apply this algorithm to study the
non-equilibrium excitonic dynamics of a chain of bacteriochlorophyll molecules,
and simulate equilibrium correlation functions of open quantum systems; finally,
we end this paper with some concluding remarks. The algorithms are implemented
as a part of the \texttt{QuantumDynamics.jl}
framework~\cite{boseQuantumDynamicsjlModularApproach2023}.

\section{Motivation \& Method}\label{sec:motivation}
Before including the thermal dissipative environment, let us start by analyzing
a discretized path integral representation of the wave function propagation for
a system defined by the Hamiltonian,
\begin{align}
    \hat{H}_0 &= \epsilon\sum_{j=1}^d\dyad{j} + \sum_{j\ge k} J(\abs{j-k})\left(\dyad{j}{k} + \dyad{k}{j}\right)
\end{align}
This is a simple Frenkel-like model for exciton or charge transport with $d$
sites. The state $\ket{k}$ represents the excitation (or charge) on the $k$th
molecule or site and every other site in the ground (or neutral) states and
spans a $d$-dimensional Hilbert space. The electronic coupling between two
states, $\ket{j}$ and $\ket{k}$, is given by some function $J$ of the distance
between the two sites. Typically, the physics of dipolar interactions imply that
$J$ would decay with distance (as $R^{-3}$, where $R$ is the distance, but we do
not make that assumption for the ensuing discussion).

If we start from an initial state $\ket{\psi(0)}$, then the time
propagated wave function is given as:
\begin{align}
    \braket{s_N}{\psi(N\Delta t)} &= \prod_{j=0}^{N-1}\sum_{s_j}\mel{s_{j+1}}{\hat{U}_\Delta}{s_j}\braket{s_0}{\psi(0)},\\
    \text{where, } \hat{U}_\Delta &= \exp\left(-i \frac{\hat{H}_0 \Delta t}{\hbar}\right).\label{eq:propagator}
\end{align}
Here, $\left\{s_j\right\}$ is the path under consideration connecting the
initial state at $s_0$ to the final state at $s_N$ in $N$ time-steps. If the
dimensionality of the Hilbert space is $d$, then this scales as
$\mathcal{O}(d^N)$ as every point in the path, $s_j$, is $d$-dimensioned. The
recently developed kink-summed path integral~\cite{makriKinkSumLongMemory2024}
groups the paths in terms of the number of kinks it possesses, where a kink is
defined as a segment where $s_j\ne s_{j+1}$. Makri has recently
showed~\cite{makriKinkSumLongMemory2024} that while the number of paths with a
particular number of kinks undergoes a maximum at $\frac{N}{2}$, the net
amplitude contributed by the set of paths with a particular number of kinks
first increases and then decreases becoming practically negligible by the time
one incorporates 8--9 kinks. So, in condensed phase simulations it may not be
necessary to include all kinks to get the correct dynamics.

To understand the complexity of a kink-summed path integral wave function
simulation, one needs to count the number of $N$-time step paths involving up to
$K$. This is given by
\begin{align}
    \sum_{k=0}^{\min(N, K)}\binom{N}{k} (d-1)^k.\label{eq:kink-num-paths}
\end{align}
If all kinks are accounted for, this binomial expansion sums to be the expected
$d^N$. The computational complexity of kink-summed path integral on including up
to $K$ kinks is a polynomial of order $K$ (for $N>K$). Surprisingly, if the
number of blips that we need to incorporate is around 8 or 9, this cost
asymptotically becomes significantly larger than the polynomial dependence of
the tensor network algorithms~\cite{strathearnEfficientNonMarkovianQuantum2018,
boseTensorNetworkRepresentation2021}.

\begin{figure}
    \centering
    \subfloat[$\Delta t=0.25$]{\includegraphics{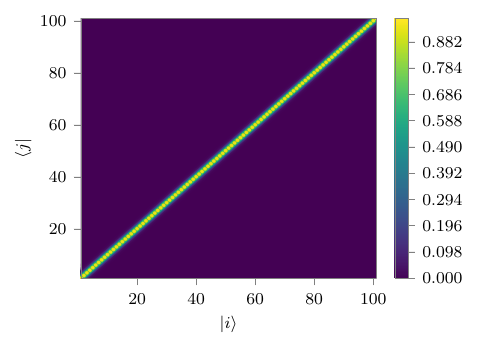}}
    
    \subfloat[$\Delta t=5.0$]{\includegraphics{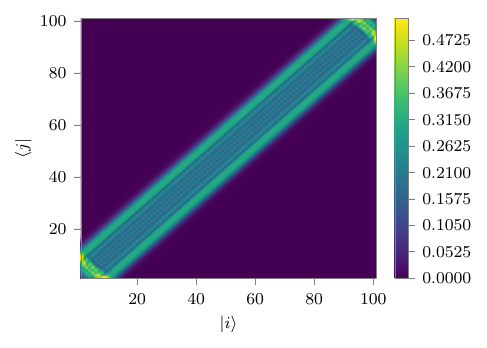}}
    \caption{Absolute value of short time propagators for the 101-mer Frenkel exciton
    model with $\epsilon = 0$, $J(R) = -\frac{1}{R^3}$ at two different values of $\Delta t$.}
    \label{fig:short-time-prop}
\end{figure}

The holy grail of algorithm design, therefore, would be to attempt to remove the
$d$-dependence of the computational complexity. To that end, we note that the
because of the relatively short-ranged interactions present in the Hamiltonian
(that is $J$ decays with distance), the propagator ends up connecting ``sites''
that are relatively close as well. In Fig.~\ref{fig:short-time-prop}, we plot
the absolute values of all the elements of the short-time propagator for a
long-distance interacting dipolar exciton-transfer model with 101 sites or
$d=101$. The propagators are calculated at $\Delta t = 0.25$ and $\Delta t =
5.0$. For the isolated system, the propagation is exact at all values of $\Delta
t$. However, as soon as we put in the environment, the system-environment
coupling would put an upper bound on the size of $\Delta t$ that is converged
due to errors associated with Trotter splitting of the propagator. Notice that
the propagator is almost completely diagonal at the shorter values of $\Delta
t$, despite having a long-range coupling $J(R) = -\frac{1}{R^3}$. This can be understood quite simply by expanding Eq.~\ref{eq:propagator},
\begin{align}
    \hat{U}_\Delta &= \mathbb{I} - i\frac{\hat{H}_0 \Delta t}{\hbar} + \mathcal{O}\left(\left(\frac{H\Delta t}{\hbar}\right)^2\right).
\end{align}
Therefore, at very short times, the structure and the sparsity of the propagator
matrix is same as that of the Hamiltonian. So, of the $d-1$ possible kink
segments, only a very few of them would contribute a non-negligible amplitude as
long as $\Delta t$ is relatively small. Even at significantly larger times, we
notice that the propagator, while not diagonal, still has a band structure and
is quite sparse. The breadth of this band is related to the maximum value of
$\abs{i-j}$ for which $\mel{i}{\hat{U}_\Delta}{j}$ has a non-negligible
contribution. This determines the number of kinks that are relevant to the
calculation.

\subsection{Path Generation Algorithm}
The goal, then, is to design an approach to path generation that leverages the
relative locality of interaction and the consequent sparsity of the propagator
matrix to generate the optimal path list. This should allow us to achieve an
asymptotic size-independence for the path integral simulations while being able
to take advantage of the latest developments that have made simulations
significantly more feasible. Consider the following algorithm of generating
paths of $N$ time-steps with a maximum of $K$ blips:
\begin{enumerate}
    \item Start with the set of all $N-1$ time-step paths with $K$ or less blips, $S^{N-1}_{K}$.
    \item For every path in $S^{N-1}_{K}$, create $N$ time-step paths by repeating the last element in the path. These would constitute the paths that do not have a kink on the final segment.
    \item For all paths in $S^{N-1}_K$ with less than $K$ kinks, create $N$ time-step paths by conditionally appending an element that is not the final element of the path. The condition is that if $l\ne p_{N-1}$ is the state that is being considered for extending the path, and $p_{N-1}$ is the final element of the path $p$ in $S^{N-1}_K$, then $\abs{\mel{l}{U_\Delta}{p_{N-1}}} \ge \chi \abs{\mel{p_{N-1}}{U_\Delta}{p_{N-1}}}$, where $\chi$ is some user-defined threshold.
\end{enumerate}
Starting with all possible paths for $N=0$ time-steps, the above three steps
give us an inductive algorithm for generating the path lists for all other $N$s
and $K$s. Notice that apart from a possible filtration by the number of kinks
($K$), only those kinks are generated the absolute value of whose amplitudes
turn out to be within a threshold of the corresponding non-kink segment, $\chi$.
This is the idea behind the adaptive kink-filtration. In terms of
Fig.~\ref{fig:short-time-prop}, we are trying to restrict our propagator to only
allow for kinks within the central band where the propagator matrix elements are
non-negligible. The value of $\chi$ is used as a convergence parameter and
systematically decreased till convergence. In a similar vein, the number of
permissible kinks would be increased till convergence. Notice that neither of
these is an \textit{ad hoc} approximation and can be systematically relaxed till
convergence. In this way we can generate an adaptive path list which reflects
the structure of the system Hamiltonian while remaining compatible with the
kink-summation ideas~\cite{makriKinkSumLongMemory2024}.

\begin{figure}
    \hspace*{-1.5em}
    \subfloat[Dynamics]{\includegraphics{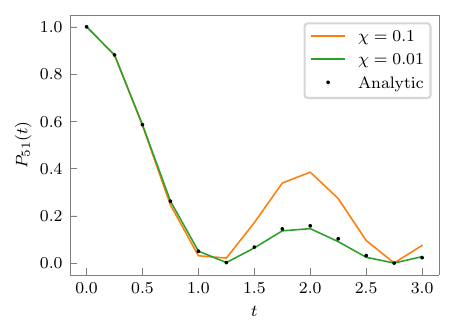}}
    
    \subfloat[Number of paths]{\includegraphics{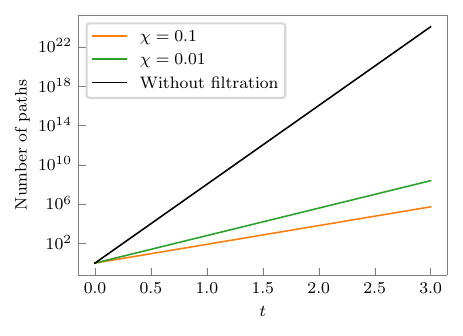}}
    \caption{Survival probability of the middle monomer of a nearest-neighbor interacting aggregate with 101 monomers at two different values of the filtration cutoff $\chi$ and the corresponding number of paths.}
    \label{fig:101-dyn-npaths}
\end{figure}

Before launching into examples of open quantum systems, first as an instructive
exercise, let us see what kind of impact this adaptive kink-filtration has on
wave-function propagation of bare excitonic systems. Consider the 101-mer long
nearest-neighbor interacting Frenkel chain, with $\epsilon=0$ and $J(R) = 1.0$
if $R=1$ and $J(R)=0.0$ otherwise, we show a comparison between the analytical
dynamics of the middle site and the number of paths involved in
Fig.~\ref{fig:101-dyn-npaths}. A cutoff of $\chi=0.01$ is sufficient in
converging the dynamics upto $\Delta t=3$. Paths with all kinks were considered
in this illustrative simulation. A na\"ive full path calculation would use
$101^N$ paths for a $N$ time-step calculation. Whereas with $\chi=0.1$, we get a
scaling of $3^N$ and at $\chi=0.01$, the scaling increases to $7^N$. It is
obvious that the filtration is extremely consequential in lowering the number of
paths. In fact, using the adaptive algorithm outlined here, the full path list
is never even generated. It should be reiterated that the goal of the method is
not to simulate isolated systems, where the results can be obtained trivially,
but to enable simulations of condensed phase systems using a scheme that allows
for systematic convergence as demonstrated in the next section. This
illustration was just provided to establish the intuition behind the scheme.

\begin{figure}
    \centering
    \includegraphics{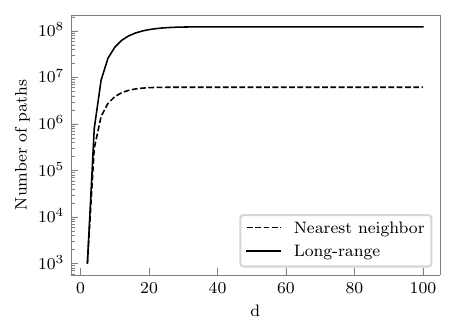}
    \caption{Number of paths versus system size, $d$, for a 8-kink calculation out to 10 time steps for a threshold of $\varepsilon=0.01$.}
    \label{fig:num_paths_size}
\end{figure}

To further investigate the effect of system size on the number of filtered kink
paths, let us consider linear aggregates of sizes from 2 (dimer) to 100 with
either a nearest neighbor coupling of $J=-1$ and the long ranged coupling of
$J(R)=-R^{-3}$. In Fig.~\ref{fig:num_paths_size}, we show the number of paths as
a function of the system size. Notice that in both cases, when filtered with
$\varepsilon=0.01$, the number of paths included in a 6-kink calculation of 10
time-step paths very quickly reaches a constant number. This constant number is
of course dependent upon the sort of interaction present in the system. For the
nearest-neighbor interacting system, the number is significantly smaller than
long-range interacting system. This is in contrast to the unfiltered 8-kink
paths that will continue to grow as $\mathcal{O}(d^8)$ irrespective of the type
of interactions present in the system. Thus, by utilizing the local nature of
the short-time propagator, one can make path integrals scale independently of
the system size for asymptotically large systems.

\section{Numerical Examples of Open Quantum System}\label{sec:numerical-examples}
Till now we explored the idea in the context of wave function dynamics of
isolated systems. Most systems of interest are not actually isolated and
interact with thermal environments. In such cases, the effect of the environment
needs to be accounted for. The Hamiltonian describing such a system-environment
coupled problem is generally expressed as follows:
\begin{align}
    \hat{H} &= \hat{H}_0 + \hat{H}_\text{sys-env},
\end{align}
where $\hat{H}_0$ is the system Hamiltonian and $\hat{H}_\text{sys-env}$
accounts for the environment and the interaction terms. Under Gaussian response,
the molecular environment can often be mapped on to one or more baths of
harmonic oscillators
\begin{align}
    \hat{H}_\text{sys-env} &= \sum_{b=1}^{N_b} \sum_{j=1}^{N_\text{osc}} \frac{p_{bj}^2}{2} + \frac{1}{2}\omega_{bj}^2\left(x_{bj} - \frac{c_{bj} \hat{s}_b}{\omega_{bj}^2}\right)^2.
\end{align}
Here, $N_b$ is assumed to be the number of baths, each with $N_\text{osc}$
oscillators. The $b$th bath interacts with the system through the operator,
$\hat{s}_b$. For simplicity, it is also assumed that the
$\comm{\hat{s}_{b_1}}{\hat{s}_{b_2}} = 0$ for any $b_1\ne b_2$. The
frequency, $\omega_{bj}$, and coupling, $c_{bj}$, of the $j$th mode of the $b$th
bath are linked via the spectral density describing this bath,
\begin{align}
    J_{b}(\omega) = \frac{\pi}{2}\sum_j \frac{c_{bj}^2}{\omega_{bj}}\delta\left(\omega-\omega_{bj}\right).
\end{align}
The spectral density is crucially important to the dynamics and can be estimated
from experiments~\cite{ratsepDemonstrationInterpretationSignificant2011} or
simulations~\cite{olbrichTheorySimulationEnvironmental2011} or a mixture
thereof. It is related to the energy-gap auto-correlation function.

\begin{figure}
    \centering
    \includegraphics{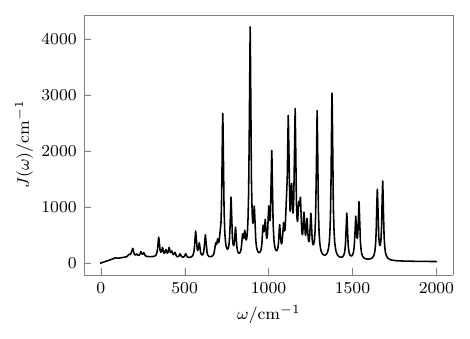}
    \caption{Spectral density for bacteriochlorophyll molecule.}
    \label{fig:spectraldensity}
\end{figure}

As an example of an open quantum system consider a chain of nearest-neighbor
interacting bacteriochlorophyll molecules. The electronic coupling is
$J=\SI{-363}{\per\cm}$. Every molecule is coupled to vibronic degrees of freedom
which are described by local harmonic baths. The spectral density, shown in
Fig.~\ref{fig:spectraldensity}, is the same as the one studied in
Ref.~\cite{sharmaImpactLossMechanisms2024} It is taken to be the sum of
the 50 relevant rigid vibrations and their Huang-Rhys factors as reported by
R\"atsep \textit{et al}.~\cite{ratsepDemonstrationInterpretationSignificant2011}
and an unstructured Brownian portion with a reorganization energy of
\SI{109}{\per\cm}. The contribution of the rigid vibrations and their Huang-Rhys
factors to the spectral density has been broadened using a Lorentzian of width
\SI{10.97}{\per\cm}.

\subsection{Non-Equilibrium Dynamics using QuAPI}
If the initial state does not have system-environment entanglement and the
environment is in a thermal state, then
$\rho(0)=\tilde\rho(0)\otimes\frac{\exp(-\beta H)}{Z_\text{env}}$, then the
dynamics of the reduced density matrix corresponding to the system is given in
the path integral formalism as:
\begin{widetext}
\begin{align}
    \mel{s_N^+}{\tilde\rho(N\Delta t)}{s_N^-} &= \sum_{s_{N-1}^\pm}\sum_{s_{N-2}^\pm}\cdots\sum_{s_0^\pm} \mel{s_N^+}{U_\Delta}{s_{N-1}^+}\mel{s_{N-1}^+}{U_\Delta}{s_{N-2}^+}\cdots\mel{s_1^+}{U_\Delta}{s_0^+}\mel{s_0^+}{\tilde\rho(0)}{s_0^-}\nonumber\\
    &\times \mel{s_0^-}{U^\dag_\Delta}{s_1^-}\mel{s_1^-}{U^\dag_\Delta}{s_2^-}\cdots\mel{s_{N-1}^-}{U^\dag_\Delta}{s_N^-}\times F\left[s^\pm(t)\right],\label{eq:real-time-pi}
\end{align}
\end{widetext}
where $s^\pm_j$ are the forward-backward path points at the $j$th time-step, $F$
is the Feynman-Vernon influence
functional~\cite{feynmanTheoryGeneralQuantum1963} and is related to the spectral
density describing the solvent~\cite{makriTensorPropagatorIterativeI1995,
makriTensorPropagatorIterativeII1995} and the energy-gap correlation
function~\cite{boseZerocostCorrectionsInfluence2022}. This influence functional
makes the dynamics non-Markovian, though in condensed phases, the length of the
memory caused by the environment is finite. By generating the forward and the
backward paths in the adaptive kink-filtered manner, we already get the benefits
of the sparse nature of the bare propagator caused by the short-ranged
interactions.

However, further enhancements can be done by filtering by the absolute magnitude
using a cutoff threshold~\cite{simTensorPropagatorWeightselected1996}, and
finally by considering forward-backward paths with less than a particular number
of blips~\cite{makriBlipDecompositionPath2014}. Blips are the time-points where
the forward path differs from the backward paths. Each blip leads to an
exponential decrease in the amplitude contributed by the path and has been
successfully used to decompose the path sum resulting in exponential speedups
for some problems~\cite{makriBlipDecompositionPath2014,
makriIterativeBlipsummedPath2017}. Here the number of blips is used as a mere
means of path filtration to additionally enhance the efficiency of the
simulations. This is done by considering only those forward and backward path
pair that differ at less than the permissible number of blips.

\begin{figure}
    \centering
    \includegraphics{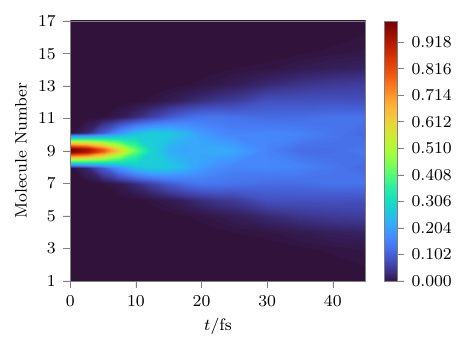}
    \caption{Excitonic dynamics in a chain of 17 bacteriochlorophyll molecules at $T=\SI{300}{\kelvin}$.}
    \label{fig:17mer}
\end{figure}

\begin{figure}
    \includegraphics{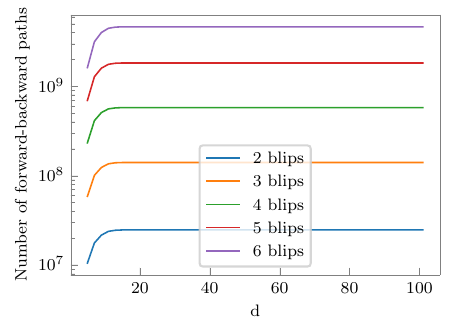}
    \caption{Number of paths at $t=\SI{50}{\fs}$ as a function of size of system, $d$, for a simulation with 8 kinks at $\Delta t=\SI{5}{\fs}$, $\chi=0.01$, and a total amplitude cutoff of $10^{-8}$.}
    \label{fig:quapi-num-paths}
\end{figure}

We demonstrate the full-memory simulation of the excitonic dynamics in a 17-mer
chain of bacteriochlorophyll molecules at $T=\SI{300}{\kelvin}$ up to
\SI{40}{\fs} using the adaptive kink filtration in Fig.~\ref{fig:17mer}. The
initial condition is taken to be $\tilde\rho(0) = \dyad{9}$, which is the middle
monomer. The simulation used time-steps of $\Delta t=\SI{2.5}{\fs}$ summing up
paths with at most 9 kinks and 10 blips till $t=\SI{25}{\fs}$, thereafter it
used a time-step of $\Delta t=\SI{5}{\fs}$. As a demonstration of the
effectiveness of the adaptive kink filtration method, the number of
forward-backward paths summed for the tenth step of the simulation at
$t=\SI{50}{\fs}$ is shown in Fig.~\ref{fig:quapi-num-paths}. Because the lines
correspond to an 8-kink simulation, according to Eq.~\ref{eq:kink-num-paths},
the number of paths should go as an 8-power polynomial of the system size, $d$.
However notice that irrespective of the maximum number of blips considered, the
number of paths saturates quite quickly. It becomes constant from around 12
monomers onward. While the exact number of paths would differ with cutoffs and
number of kinks and other factors, the trend of saturating with the number of
monomers would be invariant.

These simulations are all done at full memory. There are currently three ways of
going beyond this full memory regime: (1) traditional iteration
techniques~\cite{makriTensorPropagatorIterativeI1995,
makriTensorPropagatorIterativeII1995}; (2) the transfer tensor
method~\cite{cerrilloNonMarkovianDynamicalMaps2014}; or (3) the small matrix
decomposition~\cite{makriSmallMatrixDisentanglement2020}. Any of these methods
would require computation of the dynamical map, $\mathcal{E}(t)$, which connects
the time-evolved reduced density matrix of the system to the initial condition,
$\tilde\rho(t) = \mathcal{E}(t)\tilde\rho(0)$. For a system of size $d$, this
necessitates $d^2$ path integral calculations. Consequently, even though the
number of paths in a single path integral calculation saturates to a constant
value as $d\to\infty$, the number of paths required to obtain the dynamical map
would asymptotically go as $\mathcal{O}(d^2)$.

\subsection{Equilibrium Correlation Functions}
Finally, this adaptive kink filtration is not just limited to simulation of
non-equilibrium dynamics. It's ability to generate the most apt set of paths can
be directly utilized in improving the efficiency of calculation of correlation
functions as well. For illustration we will concentrate absorption and emission
spectra of molecular aggregates. Consider a thermal quantum correlation
function:
\begin{align}
    C_{AB}(t) \propto \Tr\left(\hat{B}(t) \hat{A}(0) \exp(-\beta \hat{H})\right)
\end{align}
where $\hat{A}$ and $\hat{B}$ are the relevant system operators. For absorption
spectrum $\hat{A} = \hat{\mu}_+$ which is the excitation operator, and $\hat{B}
= \hat{\mu}_-$ which is the de-excitation operator. However, in the case of
emission spectrum, the operators are reversed ($\hat{A}=\hat{\mu}_-$ and
$\hat{B}=\hat{\mu}_+$). The spectra are obtained as the Fourier transform of
these correlation functions.~\cite{buserInitialSystemenvironmentCorrelations2017} 
\begin{align}
    \sigma_\text{abs}(\omega) &= \int C_{\mu_+\mu_-}(t) \exp(i\omega t)\,dt\\
    \sigma_\text{ems}(\omega) &= \int C_{\mu_-\mu_+}(t) \exp(-i\omega t)\,dt
\end{align}

\begin{figure}
    \centering
    \includegraphics[scale=0.25]{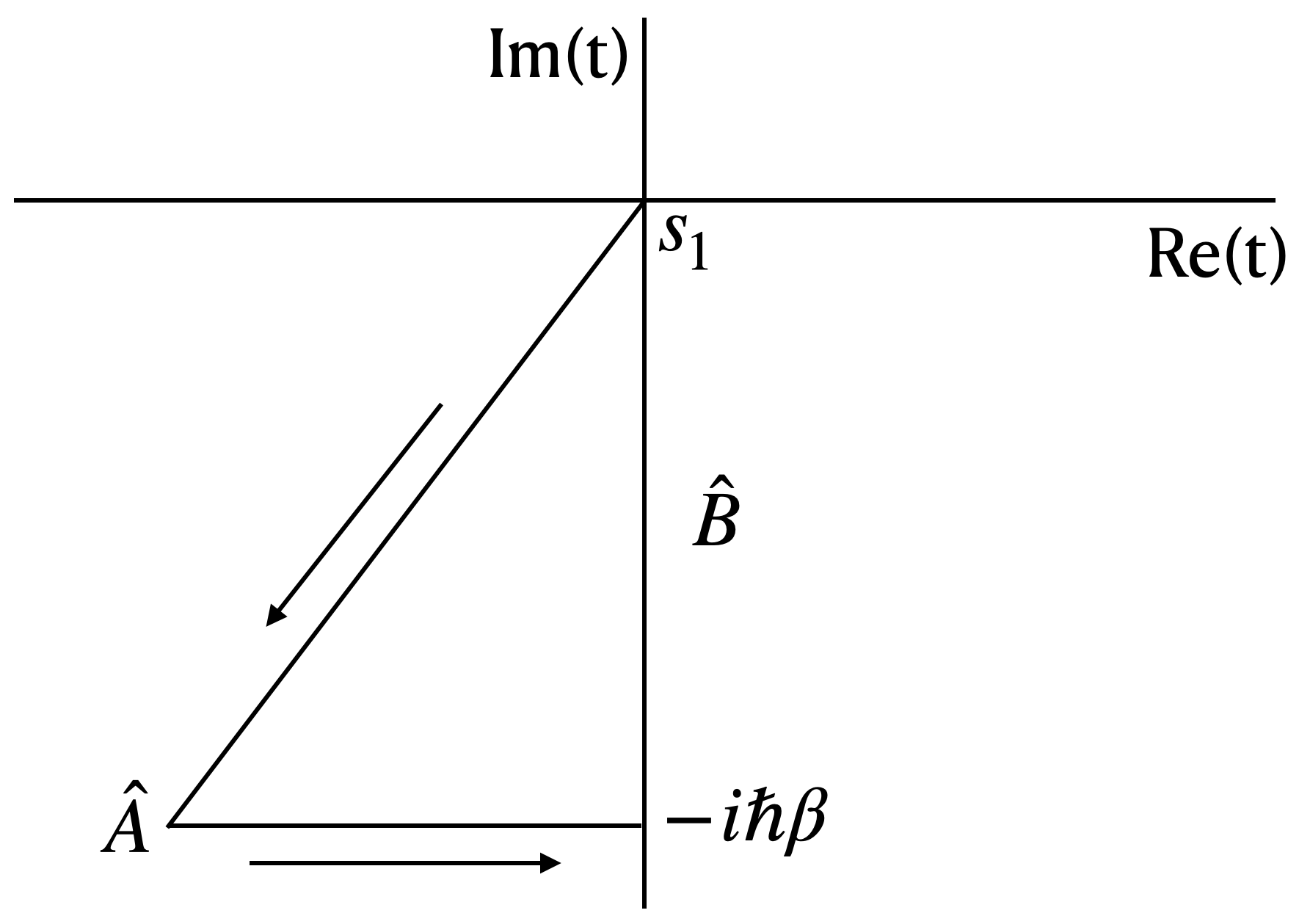}
    \caption{Complex-time contour used for calculating the absorption and emission spectra. States $s_1$ to $s_N$ are on the diagonal line, and $s_{N+1}$ to $s_{2N+2}$ are on the horizontal line connecting to the $t=-i\hbar\beta$ point.}
    \label{fig:complex-time-contour}
\end{figure}

\begin{figure}
    \centering
    \subfloat[Absorption Spectrum]{\includegraphics{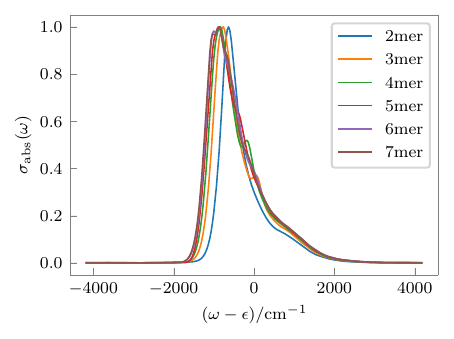}}
    
    \subfloat[Emission Spectrum]{\includegraphics{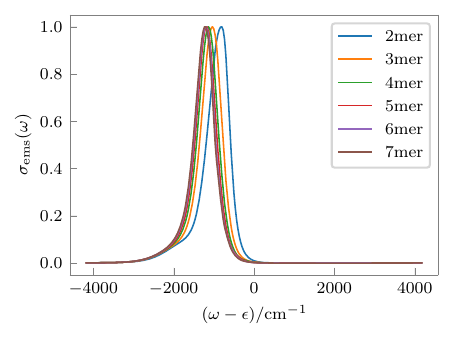}}
    \caption{Absorption and emission spectra of aggregates of different sizes.}
    \label{fig:abs-ems}
\end{figure}

Such correlation functions can be computed using the path integral
formalism~\cite{topalerQuantumRatesDouble1994,
boseQuantumCorrelationFunctions2023} to account for the environment effects.
Here we adopt a complex time contour that is obtained by representing the
correlation function in a way that combines the equilibrium density matrix with
the backward propagator:
\begin{align}
    C_{AB}(t) &\propto \Tr\left(\exp\left(\frac{i \hat{H} t_c^*}{\hbar}\right) \hat{B} \exp\left(-\frac{i \hat{H} t}{\hbar}\right) \hat{A}\right)
\end{align}
where $t_c = t - i\beta\hbar$. (Notice that this definition of the complex-time
differs from what is traditionally used when calculating symmetrized
complex-time correlation functions~\cite{topalerQuantumRatesDouble1994,
boseQuantumCorrelationFunctions2023} by a factor of 2 in the imaginary part.
This contour in complex-time is shown in Fig.~\ref{fig:complex-time-contour}.)
Consequently, the corresponding path integral becomes 
\begin{align}
    C_{AB}(t) &= \sum_{s_1}\sum_{s_2}\cdots\sum_{s_{2N+2}} \mel{s_{1}}{B}{s_{2N+2}}\nonumber\\
    &\times\mel{s_{2N+2}}{U_\Delta}{s_{2N+1}}\mel{s_{2N+1}}{U_\Delta}{s_{2N}}\ldots\nonumber\\
    &\times\mel{s_{N+2}}{U_\Delta}{s_{N+1}}\mel{s_{N+1}}{\hat{A}}{s_N}\mel{s_{N}}{U_{\Delta c}}{s_{N-1}}\nonumber\\
    &\times\ldots\times\mel{s_{3}}{U_{\Delta c}}{s_{2}}\mel{s_{2}}{U_{\Delta c}}{s_{1}}\times F_c\left[\left\{s_j\right\}\right],\\
    \text{where }U_{\Delta c} &= \exp\left(i \hat{H}_0 \frac{t}{N\hbar} - \hat{H}_0\frac{\beta}{N}\right)\\
    \text{and }U_\Delta &= \exp\left(-i\hat{H}_0\frac{t}{N\hbar}\right)
\end{align}
Here, the influence functional, $F_c$, has been derived by Topaler and
Makri~\cite{topalerQuantumRatesDouble1994} in terms of the spectral density and
has a different form in comparison to the one used for real-time dynamics in
Eq.~\ref{eq:real-time-pi}. The paths along the complex-time contour,
Fig.~\ref{fig:complex-time-contour}, are discretized as $s_j$.

In these cases, the adaptive kink procedure is trivially used to generate the
sets of half-paths, $s_1$ to $s_{N+1}$, and $s_{N+2}$ to $s_{2N+2}$ using
different short-time propagators, $U_{\Delta c}$ and $U_\Delta$ respectively.
These can then be concatenated and used for obtaining the influence functional
contributions. Because the complex-time propagators are further damped and
consequently more sparse, we expect the efficiency of the adaptive kink
filtration procedure to increase in comparison to the real time computation
counterparts.

We simulate the absorption and emission spectra of bacteriochlorophyll chains of
different sizes ranging from a dimer to a heptamer. The spectra are shown in
Fig.~\ref{fig:abs-ems}. We notice a distinct red shift of both the emission and
absorption lineshapes with increasing system size before convergence to a
particular location. The absorption lineshape is significantly broader than the
emission lineshape in all cases. Its width also increases on going from a dimer
to a trimer. Additionally, clearly visible from the the plots is the Stoke's
shift between the absorption and emission peaks, which is also size-dependent.
For a dimer it is around $\SI{167}{\per\cm}$, and it increases to around
$\SI{334}{\per\cm}$.

\section{Conclusion}
In this paper we have addressed the long-standing issue of system size
dependence of the computational complexity of path integral calculations. While
in the recent times, a lot of progress has happened in terms of reducing the
scaling of the problem in both in terms of the number of time-steps and the size
of the system, even the best of these methods scale as high powered polynomials
of the system dimensionality. Thus, even with these developments, simulations
are limited to relatively small-ish cluster sizes. We design and present an
adaptive kink filtration method that addresses this challenge, providing a first
step of moving towards even larger systems.

In most physical systems the interactions decay with distance, leading to
relatively short-ranged interactions, which induce sparsity in the system
propagator operator. We utilize this sparsity in developing an algorithm of path
generation that ensures that the number of paths considered saturates after a
certain system-dependent. The basic physical intuition is that in a large chain,
sites that are separated by large distances should not be really coupled and
consequently, the amplitude for the system to hop between the said sites would
be negligible. The adaptive kink filtration algorithm is a formalization of this
simple intuition. Thus asymptotically, the number of paths and therefore the
complexity becomes a constant with the system size.


The adaptive kink filtration method does not assume any particular structure of
the Hamiltonian as is assumed in modular path
integral~\cite{makriModularPathIntegral2018} or as is convenient for multi-site
tensor network path integral~\cite{boseMultisiteDecompositionTensor2022}. It
simply creates a built-in adaptive algorithm that automatically generates only
the most relevant paths to be added to the path list irrespective of the
underlying structure of the system. The physics of the particular system
Hamiltonian and the particular initial condition under study is automatically
taken into account. In this sense it is similar to other filtration algorithms.
Additionally, this technique is fully compatible with existing techniques such
as magnitude-based~\cite{simTensorPropagatorWeightselected1996,
simFilteredPropagatorFunctional1997} and blip-based
filtration~\cite{makriBlipDecompositionPath2014} and the kink
sum~\cite{makriKinkSumLongMemory2024} method.

We demonstrate the adaptive filtration technique by first studying the
non-equilibrium exciton dynamics in a chain of bacteriochloropyll molecules
using the Feynman-Vernon influence functional. We show the saturation of the
number of paths as the number of molecules is increased. In this example, in
addition to the adaptive kink filtration, we also use the path-based
filtration~\cite{simTensorPropagatorWeightselected1996} and blip
filtration~\cite{makriBlipDecompositionPath2014} for additional efficiency.

Finally, we explore the possibility of using the adaptive kink-filtration to
make simulations of correlation functions more efficient. To this end we
consider the absorption and emission spectra of excitonic aggregates in presence
of vibrations. In this case, the adaptive kink-filtration algorithm is used to
generate the purely real-time and the complex-time halves of the path separately
and then input into the path integral calculation. This demonstrates the
versatility of the algorithm that stems from its simplicity.

Code implementing this algorithm has already been released in the
\texttt{QuantumDynamics.jl}~\cite{boseQuantumDynamicsjlModularApproach2023}
framework. The details would be discussed in an upcoming publication detailing
the additions to the package. We believe that in conjunction with other recent
developments oriented towards improving the scaling of path integral
simulations, the adaptive kink-filtration technique outlined here would prove to
be a key step in making simulations of large systems possible. Future study
would focus on exciton-polaritonic systems amongst others because the almost
uniform coupling of the molecules to the cavity mode provides an interesting
case where the adaptive kink-filtration would prove to be super efficient in
going to large systems.

\bibliography{library}

\begin{thebibliography}{40}%
\makeatletter
\providecommand \@ifxundefined [1]{%
 \@ifx{#1\undefined}
}%
\providecommand \@ifnum [1]{%
 \ifnum #1\expandafter \@firstoftwo
 \else \expandafter \@secondoftwo
 \fi
}%
\providecommand \@ifx [1]{%
 \ifx #1\expandafter \@firstoftwo
 \else \expandafter \@secondoftwo
 \fi
}%
\providecommand \natexlab [1]{#1}%
\providecommand \enquote  [1]{``#1''}%
\providecommand \bibnamefont  [1]{#1}%
\providecommand \bibfnamefont [1]{#1}%
\providecommand \citenamefont [1]{#1}%
\providecommand \href@noop [0]{\@secondoftwo}%
\providecommand \href [0]{\begingroup \@sanitize@url \@href}%
\providecommand \@href[1]{\@@startlink{#1}\@@href}%
\providecommand \@@href[1]{\endgroup#1\@@endlink}%
\providecommand \@sanitize@url [0]{\catcode `\\12\catcode `\$12\catcode `\&12\catcode `\#12\catcode `\^12\catcode `\_12\catcode `\%12\relax}%
\providecommand \@@startlink[1]{}%
\providecommand \@@endlink[0]{}%
\providecommand \url  [0]{\begingroup\@sanitize@url \@url }%
\providecommand \@url [1]{\endgroup\@href {#1}{\urlprefix }}%
\providecommand \urlprefix  [0]{URL }%
\providecommand \Eprint [0]{\href }%
\providecommand \doibase [0]{https://doi.org/}%
\providecommand \selectlanguage [0]{\@gobble}%
\providecommand \bibinfo  [0]{\@secondoftwo}%
\providecommand \bibfield  [0]{\@secondoftwo}%
\providecommand \translation [1]{[#1]}%
\providecommand \BibitemOpen [0]{}%
\providecommand \bibitemStop [0]{}%
\providecommand \bibitemNoStop [0]{.\EOS\space}%
\providecommand \EOS [0]{\spacefactor3000\relax}%
\providecommand \BibitemShut  [1]{\csname bibitem#1\endcsname}%
\let\auto@bib@innerbib\@empty
\bibitem [{\citenamefont {White}(1992)}]{whiteDensityMatrixFormulation1992}%
  \BibitemOpen
  \bibfield  {author} {\bibinfo {author} {\bibfnamefont {S.~R.}\ \bibnamefont {White}},\ }\bibfield  {title} {\enquote {\bibinfo {title} {Density matrix formulation for quantum renormalization groups},}\ }\href {https://doi.org/10.1103/physrevlett.69.2863} {\bibfield  {journal} {\bibinfo  {journal} {Physical Review Letters}\ }\textbf {\bibinfo {volume} {69}},\ \bibinfo {pages} {2863--2866} (\bibinfo {year} {1992})}\BibitemShut {NoStop}%
\bibitem [{\citenamefont {White}\ and\ \citenamefont {Feiguin}(2004)}]{whiteRealTimeEvolutionUsing2004}%
  \BibitemOpen
  \bibfield  {author} {\bibinfo {author} {\bibfnamefont {S.~R.}\ \bibnamefont {White}}\ and\ \bibinfo {author} {\bibfnamefont {A.~E.}\ \bibnamefont {Feiguin}},\ }\bibfield  {title} {\enquote {\bibinfo {title} {Real-{{Time Evolution Using}} the {{Density Matrix Renormalization Group}}},}\ }\href {https://doi.org/10.1103/physrevlett.93.076401} {\bibfield  {journal} {\bibinfo  {journal} {Physical Review Letters}\ }\textbf {\bibinfo {volume} {93}},\ \bibinfo {pages} {076401} (\bibinfo {year} {2004})}\BibitemShut {NoStop}%
\bibitem [{\citenamefont {Paeckel}\ \emph {et~al.}(2019)\citenamefont {Paeckel}, \citenamefont {K{\"o}hler}, \citenamefont {Swoboda}, \citenamefont {Manmana}, \citenamefont {Schollw{\"o}ck},\ and\ \citenamefont {Hubig}}]{paeckelTimeevolutionMethodsMatrixproduct2019}%
  \BibitemOpen
  \bibfield  {author} {\bibinfo {author} {\bibfnamefont {S.}~\bibnamefont {Paeckel}}, \bibinfo {author} {\bibfnamefont {T.}~\bibnamefont {K{\"o}hler}}, \bibinfo {author} {\bibfnamefont {A.}~\bibnamefont {Swoboda}}, \bibinfo {author} {\bibfnamefont {S.~R.}\ \bibnamefont {Manmana}}, \bibinfo {author} {\bibfnamefont {U.}~\bibnamefont {Schollw{\"o}ck}},\ and\ \bibinfo {author} {\bibfnamefont {C.}~\bibnamefont {Hubig}},\ }\bibfield  {title} {\enquote {\bibinfo {title} {Time-evolution methods for matrix-product states},}\ }\href {https://doi.org/10.1016/j.aop.2019.167998} {\bibfield  {journal} {\bibinfo  {journal} {Annals of Physics}\ }\textbf {\bibinfo {volume} {411}},\ \bibinfo {pages} {167998} (\bibinfo {year} {2019})}\BibitemShut {NoStop}%
\bibitem [{\citenamefont {Meyer}, \citenamefont {Manthe},\ and\ \citenamefont {Cederbaum}(1990)}]{meyerMulticonfigurationalTimedependentHartree1990}%
  \BibitemOpen
  \bibfield  {author} {\bibinfo {author} {\bibfnamefont {H.-D.}\ \bibnamefont {Meyer}}, \bibinfo {author} {\bibfnamefont {U.}~\bibnamefont {Manthe}},\ and\ \bibinfo {author} {\bibfnamefont {L.}~\bibnamefont {Cederbaum}},\ }\bibfield  {title} {\enquote {\bibinfo {title} {The multi-configurational time-dependent {{Hartree}} approach},}\ }\href {https://doi.org/10.1016/0009-2614(90)87014-I} {\bibfield  {journal} {\bibinfo  {journal} {Chemical Physics Letters}\ }\textbf {\bibinfo {volume} {165}},\ \bibinfo {pages} {73--78} (\bibinfo {year} {1990})}\BibitemShut {NoStop}%
\bibitem [{\citenamefont {Wang}\ and\ \citenamefont {Thoss}(2003)}]{wangMultilayerFormulationMulticonfiguration2003}%
  \BibitemOpen
  \bibfield  {author} {\bibinfo {author} {\bibfnamefont {H.}~\bibnamefont {Wang}}\ and\ \bibinfo {author} {\bibfnamefont {M.}~\bibnamefont {Thoss}},\ }\bibfield  {title} {\enquote {\bibinfo {title} {Multilayer formulation of the multiconfiguration time-dependent {{Hartree}} theory},}\ }\href {https://doi.org/10.1063/1.1580111} {\bibfield  {journal} {\bibinfo  {journal} {The Journal of Chemical Physics}\ }\textbf {\bibinfo {volume} {119}},\ \bibinfo {pages} {1289--1299} (\bibinfo {year} {2003})}\BibitemShut {NoStop}%
\bibitem [{\citenamefont {Wang}(2015)}]{wangMultilayerMulticonfigurationTimeDependent2015}%
  \BibitemOpen
  \bibfield  {author} {\bibinfo {author} {\bibfnamefont {H.}~\bibnamefont {Wang}},\ }\bibfield  {title} {\enquote {\bibinfo {title} {Multilayer {{Multiconfiguration Time-Dependent Hartree Theory}}},}\ }\href {https://doi.org/10.1021/acs.jpca.5b03256} {\bibfield  {journal} {\bibinfo  {journal} {The Journal of Physical Chemistry A}\ }\textbf {\bibinfo {volume} {119}},\ \bibinfo {pages} {7951--7965} (\bibinfo {year} {2015})}\BibitemShut {NoStop}%
\bibitem [{\citenamefont {Feynman}\ and\ \citenamefont {Vernon}(1963)}]{feynmanTheoryGeneralQuantum1963}%
  \BibitemOpen
  \bibfield  {author} {\bibinfo {author} {\bibfnamefont {R.~P.}\ \bibnamefont {Feynman}}\ and\ \bibinfo {author} {\bibfnamefont {F.~L.}\ \bibnamefont {Vernon}},\ }\bibfield  {title} {\enquote {\bibinfo {title} {The theory of a general quantum system interacting with a linear dissipative system},}\ }\href {https://doi.org/10.1016/0003-4916(63)90068-x} {\bibfield  {journal} {\bibinfo  {journal} {Annals of Physics}\ }\textbf {\bibinfo {volume} {24}},\ \bibinfo {pages} {118--173} (\bibinfo {year} {1963})}\BibitemShut {NoStop}%
\bibitem [{\citenamefont {Makri}\ and\ \citenamefont {Makarov}(1995{\natexlab{a}})}]{makriTensorPropagatorIterativeI1995}%
  \BibitemOpen
  \bibfield  {author} {\bibinfo {author} {\bibfnamefont {N.}~\bibnamefont {Makri}}\ and\ \bibinfo {author} {\bibfnamefont {D.~E.}\ \bibnamefont {Makarov}},\ }\bibfield  {title} {\enquote {\bibinfo {title} {Tensor propagator for iterative quantum time evolution of reduced density matrices. {{I}}. {{Theory}}},}\ }\href {https://doi.org/10.1063/1.469509} {\bibfield  {journal} {\bibinfo  {journal} {The Journal of Chemical Physics}\ }\textbf {\bibinfo {volume} {102}},\ \bibinfo {pages} {4600--4610} (\bibinfo {year} {1995}{\natexlab{a}})}\BibitemShut {NoStop}%
\bibitem [{\citenamefont {Makri}\ and\ \citenamefont {Makarov}(1995{\natexlab{b}})}]{makriTensorPropagatorIterativeII1995}%
  \BibitemOpen
  \bibfield  {author} {\bibinfo {author} {\bibfnamefont {N.}~\bibnamefont {Makri}}\ and\ \bibinfo {author} {\bibfnamefont {D.~E.}\ \bibnamefont {Makarov}},\ }\bibfield  {title} {\enquote {\bibinfo {title} {Tensor propagator for iterative quantum time evolution of reduced density matrices. {{II}}. {{Numerical}} methodology},}\ }\href {https://doi.org/10.1063/1.469508} {\bibfield  {journal} {\bibinfo  {journal} {The Journal of Chemical Physics}\ }\textbf {\bibinfo {volume} {102}},\ \bibinfo {pages} {4611--4618} (\bibinfo {year} {1995}{\natexlab{b}})}\BibitemShut {NoStop}%
\bibitem [{\citenamefont {Makri}\ \emph {et~al.}(1996)\citenamefont {Makri}, \citenamefont {Sim}, \citenamefont {Makarov},\ and\ \citenamefont {Topaler}}]{makriLongtimeQuantumSimulation1996}%
  \BibitemOpen
  \bibfield  {author} {\bibinfo {author} {\bibfnamefont {N.}~\bibnamefont {Makri}}, \bibinfo {author} {\bibfnamefont {E.}~\bibnamefont {Sim}}, \bibinfo {author} {\bibfnamefont {D.~E.}\ \bibnamefont {Makarov}},\ and\ \bibinfo {author} {\bibfnamefont {M.}~\bibnamefont {Topaler}},\ }\bibfield  {title} {\enquote {\bibinfo {title} {Long-time quantum simulation of the primary charge separation in bacterial photosynthesis.}}\ }\href {https://doi.org/10.1073/pnas.93.9.3926} {\bibfield  {journal} {\bibinfo  {journal} {Proceedings of the National Academy of Sciences}\ }\textbf {\bibinfo {volume} {93}},\ \bibinfo {pages} {3926--3931} (\bibinfo {year} {1996})}\BibitemShut {NoStop}%
\bibitem [{\citenamefont {Tanimura}\ and\ \citenamefont {Wolynes}(1991)}]{tanimuraQuantumClassicalFokkerPlanck1991}%
  \BibitemOpen
  \bibfield  {author} {\bibinfo {author} {\bibfnamefont {Y.}~\bibnamefont {Tanimura}}\ and\ \bibinfo {author} {\bibfnamefont {P.~G.}\ \bibnamefont {Wolynes}},\ }\bibfield  {title} {\enquote {\bibinfo {title} {Quantum and classical {{Fokker-Planck}} equations for a {{Gaussian-Markovian}} noise bath},}\ }\href {https://doi.org/10.1103/PhysRevA.43.4131} {\bibfield  {journal} {\bibinfo  {journal} {Physical Review A}\ }\textbf {\bibinfo {volume} {43}},\ \bibinfo {pages} {4131--4142} (\bibinfo {year} {1991})}\BibitemShut {NoStop}%
\bibitem [{\citenamefont {Tanimura}(2020)}]{tanimuraNumericallyExactApproach2020}%
  \BibitemOpen
  \bibfield  {author} {\bibinfo {author} {\bibfnamefont {Y.}~\bibnamefont {Tanimura}},\ }\bibfield  {title} {\enquote {\bibinfo {title} {Numerically ``exact'' approach to open quantum dynamics: {{The}} hierarchical equations of motion ({{HEOM}})},}\ }\href {https://doi.org/10.1063/5.0011599} {\bibfield  {journal} {\bibinfo  {journal} {The Journal of Chemical Physics}\ }\textbf {\bibinfo {volume} {153}},\ \bibinfo {pages} {20901} (\bibinfo {year} {2020})}\BibitemShut {NoStop}%
\bibitem [{\citenamefont {Rahman}\ and\ \citenamefont {Kleinekath{\"o}fer}(2019)}]{rahmanChebyshevHierarchicalEquations2019}%
  \BibitemOpen
  \bibfield  {author} {\bibinfo {author} {\bibfnamefont {H.}~\bibnamefont {Rahman}}\ and\ \bibinfo {author} {\bibfnamefont {U.}~\bibnamefont {Kleinekath{\"o}fer}},\ }\bibfield  {title} {\enquote {\bibinfo {title} {Chebyshev hierarchical equations of motion for systems with arbitrary spectral densities and temperatures},}\ }\href {https://doi.org/10.1063/1.5100102} {\bibfield  {journal} {\bibinfo  {journal} {The Journal of Chemical Physics}\ }\textbf {\bibinfo {volume} {150}},\ \bibinfo {pages} {244104} (\bibinfo {year} {2019})}\BibitemShut {NoStop}%
\bibitem [{\citenamefont {Ikeda}\ and\ \citenamefont {Scholes}(2020)}]{ikedaGeneralizationHierarchicalEquations2020}%
  \BibitemOpen
  \bibfield  {author} {\bibinfo {author} {\bibfnamefont {T.}~\bibnamefont {Ikeda}}\ and\ \bibinfo {author} {\bibfnamefont {G.~D.}\ \bibnamefont {Scholes}},\ }\bibfield  {title} {\enquote {\bibinfo {title} {Generalization of the hierarchical equations of motion theory for efficient calculations with arbitrary correlation functions},}\ }\href {https://doi.org/10.1063/5.0007327} {\bibfield  {journal} {\bibinfo  {journal} {The Journal of Chemical Physics}\ }\textbf {\bibinfo {volume} {152}},\ \bibinfo {pages} {204101} (\bibinfo {year} {2020})}\BibitemShut {NoStop}%
\bibitem [{\citenamefont {Yan}, \citenamefont {Xing},\ and\ \citenamefont {Shi}(2020)}]{yanNewMethodImprove2020}%
  \BibitemOpen
  \bibfield  {author} {\bibinfo {author} {\bibfnamefont {Y.}~\bibnamefont {Yan}}, \bibinfo {author} {\bibfnamefont {T.}~\bibnamefont {Xing}},\ and\ \bibinfo {author} {\bibfnamefont {Q.}~\bibnamefont {Shi}},\ }\bibfield  {title} {\enquote {\bibinfo {title} {A new method to improve the numerical stability of the hierarchical equations of motion for discrete harmonic oscillator modes},}\ }\href {https://doi.org/10.1063/5.0027962} {\bibfield  {journal} {\bibinfo  {journal} {The Journal of Chemical Physics}\ }\textbf {\bibinfo {volume} {153}},\ \bibinfo {pages} {204109} (\bibinfo {year} {2020})}\BibitemShut {NoStop}%
\bibitem [{\citenamefont {Yan}\ \emph {et~al.}(2021)\citenamefont {Yan}, \citenamefont {Xu}, \citenamefont {Li},\ and\ \citenamefont {Shi}}]{yanEfficientPropagationHierarchical2021}%
  \BibitemOpen
  \bibfield  {author} {\bibinfo {author} {\bibfnamefont {Y.}~\bibnamefont {Yan}}, \bibinfo {author} {\bibfnamefont {M.}~\bibnamefont {Xu}}, \bibinfo {author} {\bibfnamefont {T.}~\bibnamefont {Li}},\ and\ \bibinfo {author} {\bibfnamefont {Q.}~\bibnamefont {Shi}},\ }\bibfield  {title} {\enquote {\bibinfo {title} {Efficient propagation of the hierarchical equations of motion using the {{Tucker}} and hierarchical {{Tucker}} tensors},}\ }\href {https://doi.org/10.1063/5.0050720} {\bibfield  {journal} {\bibinfo  {journal} {The Journal of Chemical Physics}\ }\textbf {\bibinfo {volume} {154}},\ \bibinfo {pages} {194104} (\bibinfo {year} {2021})}\BibitemShut {NoStop}%
\bibitem [{\citenamefont {Xu}\ \emph {et~al.}(2022)\citenamefont {Xu}, \citenamefont {Yan}, \citenamefont {Shi}, \citenamefont {Ankerhold},\ and\ \citenamefont {Stockburger}}]{xuTamingQuantumNoise2022}%
  \BibitemOpen
  \bibfield  {author} {\bibinfo {author} {\bibfnamefont {M.}~\bibnamefont {Xu}}, \bibinfo {author} {\bibfnamefont {Y.}~\bibnamefont {Yan}}, \bibinfo {author} {\bibfnamefont {Q.}~\bibnamefont {Shi}}, \bibinfo {author} {\bibfnamefont {J.}~\bibnamefont {Ankerhold}},\ and\ \bibinfo {author} {\bibfnamefont {J.~T.}\ \bibnamefont {Stockburger}},\ }\bibfield  {title} {\enquote {\bibinfo {title} {Taming {{Quantum Noise}} for {{Efficient Low Temperature Simulations}} of {{Open Quantum Systems}}},}\ }\href {https://doi.org/10.1103/PhysRevLett.129.230601} {\bibfield  {journal} {\bibinfo  {journal} {Physical Review Letters}\ }\textbf {\bibinfo {volume} {129}},\ \bibinfo {pages} {230601} (\bibinfo {year} {2022})}\BibitemShut {NoStop}%
\bibitem [{\citenamefont {Ke}(2023)}]{keTreeTensorNetwork2023}%
  \BibitemOpen
  \bibfield  {author} {\bibinfo {author} {\bibfnamefont {Y.}~\bibnamefont {Ke}},\ }\bibfield  {title} {\enquote {\bibinfo {title} {Tree tensor network state approach for solving hierarchical equations of motion},}\ }\href {https://doi.org/10.1063/5.0153870} {\bibfield  {journal} {\bibinfo  {journal} {The Journal of Chemical Physics}\ }\textbf {\bibinfo {volume} {158}},\ \bibinfo {pages} {211102} (\bibinfo {year} {2023})}\BibitemShut {NoStop}%
\bibitem [{\citenamefont {Cerrillo}\ and\ \citenamefont {Cao}(2014)}]{cerrilloNonMarkovianDynamicalMaps2014}%
  \BibitemOpen
  \bibfield  {author} {\bibinfo {author} {\bibfnamefont {J.}~\bibnamefont {Cerrillo}}\ and\ \bibinfo {author} {\bibfnamefont {J.}~\bibnamefont {Cao}},\ }\bibfield  {title} {\enquote {\bibinfo {title} {Non-{{Markovian Dynamical Maps}}: {{Numerical Processing}} of {{Open Quantum Trajectories}}},}\ }\href {https://doi.org/10.1103/PhysRevLett.112.110401} {\bibfield  {journal} {\bibinfo  {journal} {Physical Review Letters}\ }\textbf {\bibinfo {volume} {112}},\ \bibinfo {pages} {110401} (\bibinfo {year} {2014})}\BibitemShut {NoStop}%
\bibitem [{\citenamefont {Makri}(2020)}]{makriSmallMatrixDisentanglement2020}%
  \BibitemOpen
  \bibfield  {author} {\bibinfo {author} {\bibfnamefont {N.}~\bibnamefont {Makri}},\ }\bibfield  {title} {\enquote {\bibinfo {title} {Small matrix disentanglement of the path integral: {{Overcoming}} the exponential tensor scaling with memory length},}\ }\href {https://doi.org/10.1063/1.5139473} {\bibfield  {journal} {\bibinfo  {journal} {The Journal of Chemical Physics}\ }\textbf {\bibinfo {volume} {152}},\ \bibinfo {pages} {41104} (\bibinfo {year} {2020})}\BibitemShut {NoStop}%
\bibitem [{\citenamefont {Makri}(2014)}]{makriBlipDecompositionPath2014}%
  \BibitemOpen
  \bibfield  {author} {\bibinfo {author} {\bibfnamefont {N.}~\bibnamefont {Makri}},\ }\bibfield  {title} {\enquote {\bibinfo {title} {Blip decomposition of the path integral: {{Exponential}} acceleration of real-time calculations on quantum dissipative systems},}\ }\href {https://doi.org/10.1063/1.4896736} {\bibfield  {journal} {\bibinfo  {journal} {The Journal of Chemical Physics}\ }\textbf {\bibinfo {volume} {141}},\ \bibinfo {pages} {134117} (\bibinfo {year} {2014})}\BibitemShut {NoStop}%
\bibitem [{\citenamefont {Makri}(2017)}]{makriIterativeBlipsummedPath2017}%
  \BibitemOpen
  \bibfield  {author} {\bibinfo {author} {\bibfnamefont {N.}~\bibnamefont {Makri}},\ }\bibfield  {title} {\enquote {\bibinfo {title} {Iterative blip-summed path integral for quantum dynamics in strongly dissipative environments},}\ }\href {https://doi.org/10.1063/1.4979197} {\bibfield  {journal} {\bibinfo  {journal} {The Journal of Chemical Physics}\ }\textbf {\bibinfo {volume} {146}},\ \bibinfo {pages} {134101} (\bibinfo {year} {2017})}\BibitemShut {NoStop}%
\bibitem [{\citenamefont {Strathearn}\ \emph {et~al.}(2018)\citenamefont {Strathearn}, \citenamefont {Kirton}, \citenamefont {Kilda}, \citenamefont {Keeling},\ and\ \citenamefont {Lovett}}]{strathearnEfficientNonMarkovianQuantum2018}%
  \BibitemOpen
  \bibfield  {author} {\bibinfo {author} {\bibfnamefont {A.}~\bibnamefont {Strathearn}}, \bibinfo {author} {\bibfnamefont {P.}~\bibnamefont {Kirton}}, \bibinfo {author} {\bibfnamefont {D.}~\bibnamefont {Kilda}}, \bibinfo {author} {\bibfnamefont {J.}~\bibnamefont {Keeling}},\ and\ \bibinfo {author} {\bibfnamefont {B.~W.}\ \bibnamefont {Lovett}},\ }\bibfield  {title} {\enquote {\bibinfo {title} {Efficient non-{{Markovian}} quantum dynamics using time-evolving matrix product operators},}\ }\href {https://doi.org/10.1038/s41467-018-05617-3} {\bibfield  {journal} {\bibinfo  {journal} {Nature Communications}\ }\textbf {\bibinfo {volume} {9}},\ \bibinfo {pages} {3322} (\bibinfo {year} {2018})}\BibitemShut {NoStop}%
\bibitem [{\citenamefont {J{\o}rgensen}\ and\ \citenamefont {Pollock}(2019)}]{jorgensenExploitingCausalTensor2019}%
  \BibitemOpen
  \bibfield  {author} {\bibinfo {author} {\bibfnamefont {M.~R.}\ \bibnamefont {J{\o}rgensen}}\ and\ \bibinfo {author} {\bibfnamefont {F.~A.}\ \bibnamefont {Pollock}},\ }\bibfield  {title} {\enquote {\bibinfo {title} {Exploiting the {{Causal Tensor Network Structure}} of {{Quantum Processes}} to {{Efficiently Simulate Non-Markovian Path Integrals}}},}\ }\href {https://doi.org/10.1103/physrevlett.123.240602} {\bibfield  {journal} {\bibinfo  {journal} {Physical Review Letters}\ }\textbf {\bibinfo {volume} {123}},\ \bibinfo {pages} {240602} (\bibinfo {year} {2019})}\BibitemShut {NoStop}%
\bibitem [{\citenamefont {Bose}(2022{\natexlab{a}})}]{bosePairwiseConnectedTensor2022}%
  \BibitemOpen
  \bibfield  {author} {\bibinfo {author} {\bibfnamefont {A.}~\bibnamefont {Bose}},\ }\bibfield  {title} {\enquote {\bibinfo {title} {Pairwise connected tensor network representation of path integrals},}\ }\href {https://doi.org/10.1103/PhysRevB.105.024309} {\bibfield  {journal} {\bibinfo  {journal} {Physical Review B}\ }\textbf {\bibinfo {volume} {105}},\ \bibinfo {pages} {024309} (\bibinfo {year} {2022}{\natexlab{a}})}\BibitemShut {NoStop}%
\bibitem [{\citenamefont {Bose}(2023{\natexlab{a}})}]{boseQuantumCorrelationFunctions2023}%
  \BibitemOpen
  \bibfield  {author} {\bibinfo {author} {\bibfnamefont {A.}~\bibnamefont {Bose}},\ }\bibfield  {title} {\enquote {\bibinfo {title} {Quantum correlation functions through tensor network path integral},}\ }\href {https://doi.org/10.1063/5.0174338} {\bibfield  {journal} {\bibinfo  {journal} {The Journal of Chemical Physics}\ }\textbf {\bibinfo {volume} {159}},\ \bibinfo {pages} {214110} (\bibinfo {year} {2023}{\natexlab{a}})}\BibitemShut {NoStop}%
\bibitem [{\citenamefont {Bose}\ and\ \citenamefont {Walters}(2022{\natexlab{a}})}]{boseMultisiteDecompositionTensor2022}%
  \BibitemOpen
  \bibfield  {author} {\bibinfo {author} {\bibfnamefont {A.}~\bibnamefont {Bose}}\ and\ \bibinfo {author} {\bibfnamefont {P.~L.}\ \bibnamefont {Walters}},\ }\bibfield  {title} {\enquote {\bibinfo {title} {A multisite decomposition of the tensor network path integrals},}\ }\href {https://doi.org/10.1063/5.0073234} {\bibfield  {journal} {\bibinfo  {journal} {The Journal of Chemical Physics}\ }\textbf {\bibinfo {volume} {156}},\ \bibinfo {pages} {024101} (\bibinfo {year} {2022}{\natexlab{a}})}\BibitemShut {NoStop}%
\bibitem [{\citenamefont {Bose}\ and\ \citenamefont {Walters}(2022{\natexlab{b}})}]{boseTensorNetworkPath2022}%
  \BibitemOpen
  \bibfield  {author} {\bibinfo {author} {\bibfnamefont {A.}~\bibnamefont {Bose}}\ and\ \bibinfo {author} {\bibfnamefont {P.~L.}\ \bibnamefont {Walters}},\ }\bibfield  {title} {\enquote {\bibinfo {title} {Tensor {{Network Path Integral Study}} of {{Dynamics}} in {{B850 LH2 Ring}} with {{Atomistically Derived Vibrations}}},}\ }\href {https://doi.org/10.1021/acs.jctc.2c00163} {\bibfield  {journal} {\bibinfo  {journal} {Journal of Chemical Theory and Computation}\ }\textbf {\bibinfo {volume} {18}},\ \bibinfo {pages} {4095--4108} (\bibinfo {year} {2022}{\natexlab{b}})}\BibitemShut {NoStop}%
\bibitem [{\citenamefont {Makri}(2024)}]{makriKinkSumLongMemory2024}%
  \BibitemOpen
  \bibfield  {author} {\bibinfo {author} {\bibfnamefont {N.}~\bibnamefont {Makri}},\ }\bibfield  {title} {\enquote {\bibinfo {title} {Kink {{Sum}} for {{Long-Memory Small Matrix Path Integral Dynamics}}},}\ }\href {https://doi.org/10.1021/acs.jpcb.3c08282} {\bibfield  {journal} {\bibinfo  {journal} {The Journal of Physical Chemistry B}\ }\textbf {\bibinfo {volume} {128}},\ \bibinfo {pages} {2469--2480} (\bibinfo {year} {2024})}\BibitemShut {NoStop}%
\bibitem [{\citenamefont {Makri}(2018)}]{makriModularPathIntegral2018}%
  \BibitemOpen
  \bibfield  {author} {\bibinfo {author} {\bibfnamefont {N.}~\bibnamefont {Makri}},\ }\bibfield  {title} {\enquote {\bibinfo {title} {Modular path integral methodology for real-time quantum dynamics},}\ }\href {https://doi.org/10.1063/1.5058223} {\bibfield  {journal} {\bibinfo  {journal} {The Journal of Chemical Physics}\ }\textbf {\bibinfo {volume} {149}},\ \bibinfo {pages} {214108} (\bibinfo {year} {2018})}\BibitemShut {NoStop}%
\bibitem [{\citenamefont {Bose}(2023{\natexlab{b}})}]{boseQuantumDynamicsjlModularApproach2023}%
  \BibitemOpen
  \bibfield  {author} {\bibinfo {author} {\bibfnamefont {A.}~\bibnamefont {Bose}},\ }\bibfield  {title} {\enquote {\bibinfo {title} {{{QuantumDynamics}}.jl: {{A}} modular approach to simulations of dynamics of open quantum systems},}\ }\href {https://doi.org/10.1063/5.0151483} {\bibfield  {journal} {\bibinfo  {journal} {The Journal of Chemical Physics}\ }\textbf {\bibinfo {volume} {158}},\ \bibinfo {pages} {204113} (\bibinfo {year} {2023}{\natexlab{b}})}\BibitemShut {NoStop}%
\bibitem [{\citenamefont {Bose}\ and\ \citenamefont {Walters}(2021)}]{boseTensorNetworkRepresentation2021}%
  \BibitemOpen
  \bibfield  {author} {\bibinfo {author} {\bibfnamefont {A.}~\bibnamefont {Bose}}\ and\ \bibinfo {author} {\bibfnamefont {P.~L.}\ \bibnamefont {Walters}},\ }\bibfield  {title} {\enquote {\bibinfo {title} {A tensor network representation of path integrals: {{Implementation}} and analysis},}\ }\href@noop {} {\bibfield  {journal} {\bibinfo  {journal} {arXiv pre-print server arXiv:2106.12523}\ } (\bibinfo {year} {2021})},\ \Eprint {https://arxiv.org/abs/2106.12523} {arXiv:2106.12523} \BibitemShut {NoStop}%
\bibitem [{\citenamefont {R{\"a}tsep}\ \emph {et~al.}(2011)\citenamefont {R{\"a}tsep}, \citenamefont {Cai}, \citenamefont {Reimers},\ and\ \citenamefont {Freiberg}}]{ratsepDemonstrationInterpretationSignificant2011}%
  \BibitemOpen
  \bibfield  {author} {\bibinfo {author} {\bibfnamefont {M.}~\bibnamefont {R{\"a}tsep}}, \bibinfo {author} {\bibfnamefont {Z.-L.}\ \bibnamefont {Cai}}, \bibinfo {author} {\bibfnamefont {J.~R.}\ \bibnamefont {Reimers}},\ and\ \bibinfo {author} {\bibfnamefont {A.}~\bibnamefont {Freiberg}},\ }\bibfield  {title} {\enquote {\bibinfo {title} {Demonstration and interpretation of significant asymmetry in the low-resolution and high-resolution {{Qy}} fluorescence and absorption spectra of bacteriochlorophyll a},}\ }\href {https://doi.org/10.1063/1.3518685} {\bibfield  {journal} {\bibinfo  {journal} {The Journal of Chemical Physics}\ }\textbf {\bibinfo {volume} {134}},\ \bibinfo {pages} {24506} (\bibinfo {year} {2011})}\BibitemShut {NoStop}%
\bibitem [{\citenamefont {Olbrich}\ \emph {et~al.}(2011)\citenamefont {Olbrich}, \citenamefont {Str{\"u}mpfer}, \citenamefont {Schulten},\ and\ \citenamefont {Kleinekath{\"o}fer}}]{olbrichTheorySimulationEnvironmental2011}%
  \BibitemOpen
  \bibfield  {author} {\bibinfo {author} {\bibfnamefont {C.}~\bibnamefont {Olbrich}}, \bibinfo {author} {\bibfnamefont {J.}~\bibnamefont {Str{\"u}mpfer}}, \bibinfo {author} {\bibfnamefont {K.}~\bibnamefont {Schulten}},\ and\ \bibinfo {author} {\bibfnamefont {U.}~\bibnamefont {Kleinekath{\"o}fer}},\ }\bibfield  {title} {\enquote {\bibinfo {title} {Theory and {{Simulation}} of the {{Environmental Effects}} on {{FMO Electronic Transitions}}},}\ }\href {https://doi.org/10.1021/jz2007676} {\bibfield  {journal} {\bibinfo  {journal} {The Journal of Physical Chemistry Letters}\ }\textbf {\bibinfo {volume} {2}},\ \bibinfo {pages} {1771--1776} (\bibinfo {year} {2011})}\BibitemShut {NoStop}%
\bibitem [{\citenamefont {Sharma}\ and\ \citenamefont {Bose}(2024)}]{sharmaImpactLossMechanisms2024}%
  \BibitemOpen
  \bibfield  {author} {\bibinfo {author} {\bibfnamefont {D.}~\bibnamefont {Sharma}}\ and\ \bibinfo {author} {\bibfnamefont {A.}~\bibnamefont {Bose}},\ }\bibfield  {title} {\enquote {\bibinfo {title} {Impact of {{Loss Mechanisms}} on {{Linear Spectra}} of {{Excitonic}} and {{Polaritonic Aggregates}}},}\ }\href {https://doi.org/10.1021/acs.jctc.4c00825} {\bibfield  {journal} {\bibinfo  {journal} {Journal of Chemical Theory and Computation}\ }\textbf {\bibinfo {volume} {20}},\ \bibinfo {pages} {9522--9532} (\bibinfo {year} {2024})}\BibitemShut {NoStop}%
\bibitem [{\citenamefont {Bose}(2022{\natexlab{b}})}]{boseZerocostCorrectionsInfluence2022}%
  \BibitemOpen
  \bibfield  {author} {\bibinfo {author} {\bibfnamefont {A.}~\bibnamefont {Bose}},\ }\bibfield  {title} {\enquote {\bibinfo {title} {Zero-cost corrections to influence functional coefficients from bath response functions},}\ }\href {https://doi.org/10.1063/5.0101396} {\bibfield  {journal} {\bibinfo  {journal} {The Journal of Chemical Physics}\ }\textbf {\bibinfo {volume} {157}},\ \bibinfo {pages} {054107} (\bibinfo {year} {2022}{\natexlab{b}})}\BibitemShut {NoStop}%
\bibitem [{\citenamefont {Sim}\ and\ \citenamefont {Makri}(1996)}]{simTensorPropagatorWeightselected1996}%
  \BibitemOpen
  \bibfield  {author} {\bibinfo {author} {\bibfnamefont {E.}~\bibnamefont {Sim}}\ and\ \bibinfo {author} {\bibfnamefont {N.}~\bibnamefont {Makri}},\ }\bibfield  {title} {\enquote {\bibinfo {title} {Tensor propagator with weight-selected paths for quantum dissipative dynamics with long-memory kernels},}\ }\href {https://doi.org/10.1016/0009-2614(95)01374-1} {\bibfield  {journal} {\bibinfo  {journal} {Chemical Physics Letters}\ }\textbf {\bibinfo {volume} {249}},\ \bibinfo {pages} {224--230} (\bibinfo {year} {1996})}\BibitemShut {NoStop}%
\bibitem [{\citenamefont {Buser}\ \emph {et~al.}(2017)\citenamefont {Buser}, \citenamefont {Cerrillo}, \citenamefont {Schaller},\ and\ \citenamefont {Cao}}]{buserInitialSystemenvironmentCorrelations2017}%
  \BibitemOpen
  \bibfield  {author} {\bibinfo {author} {\bibfnamefont {M.}~\bibnamefont {Buser}}, \bibinfo {author} {\bibfnamefont {J.}~\bibnamefont {Cerrillo}}, \bibinfo {author} {\bibfnamefont {G.}~\bibnamefont {Schaller}},\ and\ \bibinfo {author} {\bibfnamefont {J.}~\bibnamefont {Cao}},\ }\bibfield  {title} {\enquote {\bibinfo {title} {Initial system-environment correlations via the transfer-tensor method},}\ }\href {https://doi.org/10.1103/PhysRevA.96.062122} {\bibfield  {journal} {\bibinfo  {journal} {Physical Review A: Atomic, Molecular, and Optical Physics}\ }\textbf {\bibinfo {volume} {96}},\ \bibinfo {pages} {062122} (\bibinfo {year} {2017})}\BibitemShut {NoStop}%
\bibitem [{\citenamefont {Topaler}\ and\ \citenamefont {Makri}(1994)}]{topalerQuantumRatesDouble1994}%
  \BibitemOpen
  \bibfield  {author} {\bibinfo {author} {\bibfnamefont {M.}~\bibnamefont {Topaler}}\ and\ \bibinfo {author} {\bibfnamefont {N.}~\bibnamefont {Makri}},\ }\bibfield  {title} {\enquote {\bibinfo {title} {Quantum rates for a double well coupled to a dissipative bath: {{Accurate}} path integral results and comparison with approximate theories},}\ }\href {https://doi.org/10.1063/1.468244} {\bibfield  {journal} {\bibinfo  {journal} {The Journal of Chemical Physics}\ }\textbf {\bibinfo {volume} {101}},\ \bibinfo {pages} {7500--7519} (\bibinfo {year} {1994})}\BibitemShut {NoStop}%
\bibitem [{\citenamefont {Sim}\ and\ \citenamefont {Makri}(1997)}]{simFilteredPropagatorFunctional1997}%
  \BibitemOpen
  \bibfield  {author} {\bibinfo {author} {\bibfnamefont {E.}~\bibnamefont {Sim}}\ and\ \bibinfo {author} {\bibfnamefont {N.}~\bibnamefont {Makri}},\ }\bibfield  {title} {\enquote {\bibinfo {title} {Filtered propagator functional for iterative dynamics of quantum dissipative systems},}\ }\href {https://doi.org/10.1016/S0010-4655(96)00130-0} {\bibfield  {journal} {\bibinfo  {journal} {Computer Physics Communications}\ }\textbf {\bibinfo {volume} {99}},\ \bibinfo {pages} {335--354} (\bibinfo {year} {1997})}\BibitemShut {NoStop}%
\end{thebibliography}%
\end{document}